\documentclass[aps,prl,twocolumn,groupedaddress,showpacs]{revtex4}
\usepackage{graphicx}
\usepackage{amsfonts}		
\usepackage{amsmath}
\usepackage{color}
\definecolor{c1}{rgb}{1, 0, 0}
\definecolor{c2}{rgb}{0, 1, 0}
\definecolor{c3}{rgb}{0, 0, 1}
\definecolor{c4}{rgb}{1, 0, 1}
\definecolor{c5}{rgb}{0, 1, 1}

\def\ie{\hbox{\it i.e. }}
\def\etal{\hbox{\it et al. }}

\def\beq{\begin{equation}}
\def\eeq{\end{equation}}
\def\bea{\begin{eqnarray}}
\def\eea{\end{eqnarray}}
\pacs{64.60.F-,05.70.Jk,64.60.ae,75.10.Hk}

\begin{document}
\title{Frozen into stripes: fate of the critical Ising model after a quench}
\author{T. Blanchard$^1$ and M. Picco$^1$}
\affiliation{$^1$CNRS, LPTHE, Universit\'e Pierre et Marie Curie,
                   UMR 7589, 4 place Jussieu, 75252 Paris cedex 05, France}

\date{\today}

\begin{abstract}
\noindent In this work we study numerically the final state of the two
dimensional ferromagnetic critical Ising model after a quench to zero
temperature.  Beginning from equilibrium at $T_c$, the system can be blocked in
a variety of infinitely long lived stripe states in addition to the ground
state.  Similar results have already been obtained for an infinite temperature
initial condition and an interesting connection to exact percolation crossing
probabilities has emerged.  Here we complete this picture by providing a new
example of stripe states precisely related to initial crossing probabilities
for various boundary conditions.  We thus show that this is not specific to
percolation but rather that it depends on the properties of spanning clusters
in the initial state. 
\end{abstract}

\maketitle

\textit{Introduction} ---
Somewhat surprisingly, even with such a simple set up as the two dimensional nearest neighbour ferromagnetic Ising model
quenched to zero temperature, a variety of final states can be reached with non trivial probabilities. It was first
noticed by Lipowski~\cite{lipowski_anomalous_1999} that an anomalous scaling for the equilibration time of the kinetic
Ising model arises from the existence of stripe states. In an interesting series of papers, Krapivsky, Redner and
collaborators noticed as well the presence of stripes states after a quench from infinite temperature, and analysed both
the dynamics and the final states.  They measured numerically the probability of getting stuck in stripe states and
found that it was around $1/3$ but were at first unable to explain it~\cite{spirin_freezing_2001,spirin_fate_2001}. In
the same situation but with different goals, Sicilia \etal studied the geometry of spin clusters, and noticed that
after a few Monte-Carlo steps at $T<T_c$, an infinite temperature system on a square lattice is very similar to the
critical percolation point for the spin clusters \cite{arenzon_exact_2007,sicilia_domain_2007}. So, even though the
initial occupation probability of $1/2$ for a type of spins is inferior to the critical site percolation probability for
the square lattice $p_c\simeq 0.59$, the system is rapidly very similar to a critical percolation system. 
This correspondence enabled Krapivsky and collaborators to give an explanation for the probabilities of appearance of stripe states in
terms of critical percolation probabilities, be it for free or periodic boundary
conditions~\cite{barros_freezing_2009,olejarz_fate_2012}. The study of these quantities has a long history starting with
Cardy and his eponymous formula~\cite{cardy_critical_1992}. This formula gives the probability of the existence of an
incipient spanning cluster at the critical percolation point in terms of hypergeometric functions. Since then a number
of percolation probabilities have been studied.  One aspect that is particularly interesting is that although Cardy and
others found those formulas using conformal field theory (CFT), \ie in a non rigorous way mathematically speaking,
mathematicians have developed rigorous tools to tackle those systems. Schramm introduced the stochastic Loewner
evolution (SLE)~\cite{schramm_scaling_2000} which describes numerous physically occurring curves, and this lead to many
results (see~\cite{lawler_scaling_2002} with Lawler and Werner e.g.). Around the same time, Smirnov proved rigorously Cardy formula and the
conformality of the critical percolation point for site percolation on the triangular
lattice~\cite{smirnov_critical_2001}.

The correspondence discovered by Krapivsky \etal is thus very interesting since it relates a non-equilibrium situation
to exact results. It is remarkably accurate and based on the observation of Sicilia \etal and general considerations on
coarsening dynamics and it was supported by numerical results on a lattice with different aspect ratios.  The underlying
question to those studies is to what extent the initial condition influences the final one?  In this regard, it is
interesting to study the same situation with a different initial condition to extract the universal behaviors from the
rest. For the Ising model, the results discussed above have to hold for any starting temperature above the critical temperature since for an initial
condition at $T>T_c$ the long distance properties are governed by the infinite temperature fixed point. Since the
subcritical region is trivial from the spin clusters point of view, the critical temperature point only remains and is
absolutely non trivial concerning spin clusters as we will disscuss below.  Actually, the persistence of the initial
condition is also very interesting to study before reaching the final state, \ie during the equilibration.  Several
works have dealt with the issue of cluster dynamics after a quench either to
$T<T_c$~\cite{arenzon_exact_2007,sicilia_domain_2007} or to $T_c$~\cite{blanchard_morphological_2012} for the Ising
model and for the Potts model~\cite{loureiro_curvature-driven_2010,loureiro_geometrical_2012} and were able to clarify
the influence of the initial condition. Indeed those works show that the initial properties of spin clusters are retained at distances
bigger than a dynamical length-scale $\xi(t)\sim t^{1/z}$ using dynamical scaling arguments and numerical checks.

In the following we study the ferromagnetic Ising model whose hamiltonian is written as:
\begin{equation}
{\cal H} = -  \sum_{\langle ij\rangle} J S_i S_j   \; ,
\end{equation}
where the sum is over all nearest neighbors pairs of spins of a two dimensional system and $J>0$. The system considered is
rectangular and three boundary conditions will be considered : free, periodic and fixed. The lattices used will be
discussed in each case.

\begin{figure}
\begin{center}
\includegraphics[scale=1.15]{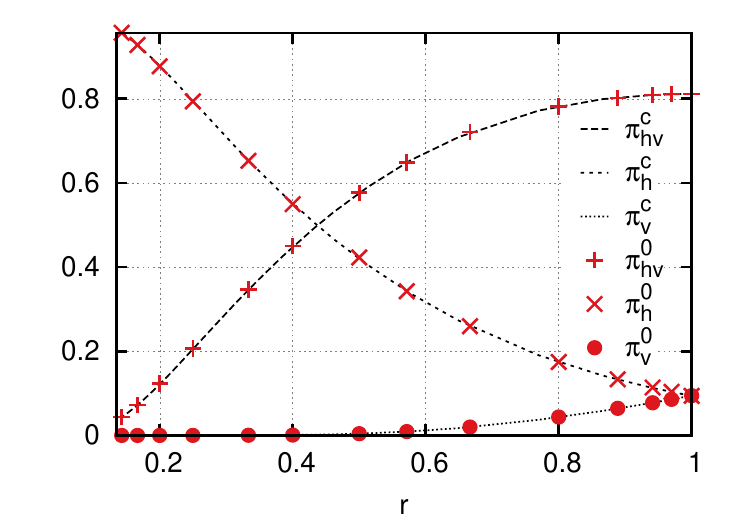}
\end{center}
\caption{
(color online) Crossing probabilities for free boundary conditions versus $r$. The lines corresponding to $\pi^c_h,
\pi^c_v$ and $\pi^c_{hv}$ are from the results given in \cite{LSA} for the
critical Ising model. The points correspond to our data for $\pi^0_h,
\pi^0_v$ and $\pi^0_{hv}$ obtained after a quench at zero temperature.}
\label{PCI}
\end{figure}

\textit{Method} ---
In the simulation that we will consider here, we started from a system
equilibrated at the critical temperature. The equilibration is obtained with
the Wolff cluster algorithm \cite{Wolff}. We generate a large number of
independent configurations, at least one million for each size considered. For
each configuration, we determine the probability that there exists at least one
cluster of spins crossing the horizontal or vertical direction.
We denote these probabilities  $\pi^c_h$ or $\pi^c_v$. The probability that a
spin cluster is crossing in both directions will be denoted $\pi^{c}_{hv}$.
Next, we perform a quench at $T=0$. We then let the system evolve with a Glauber dynamics 
with a suitable algorithm described below until it reaches a stable state. We record
similar (stripes) probabilities that we denote by $\pi^0_h, \pi^0_v$ for the
horizontal or vertical stripes and $\pi^0_{hv}$ for the stripes in both
directions, \ie a ground state.  

One difficulty is that it can take a tremendous amount of (Monte-Carlo) time to
reach the final state where no spins can be updated without increasing the
energy of the system because the system tends to wander in long-lived
iso-energetic states. A simple kinetic Monte Carlo
algorithm~\cite{bortz_new_1975} with Glauber dynamics bypasses this issue quite easily. It
essentially consists in an efficient sampling of the updateable spins which
greatly reduces computation time to reach a final state.  In a $T=0$ dynamics,
the only possible transition rates on even-coordinated lattices are $0$, $0.5$ and $1$ respectively for
spins with strictly more than half of their neighbours of the same colour as
theirs, exactly half and strictly less than half. Let us call $n_+$, $n_0$ and
$n_-$ the number of spins in this three categories and $R$ the sum of all
possible rates, with obviously $R\equiv n_-+n_0/2$. A list of the $n_-+n_0$
updateable sites is created. Now, to update the system, a random number
$u\in]0,1]$ is drawn, and if $uR\le n_-$, one of the $n_-$ spins is flipped and
otherwise one among the $n_0$ ones. To update the time $t$, we draw another
random number $u'\in]0,1]$ and $t\leftarrow t+\ln(1/u')/R$. Then the list of
updateable sites is regenerated and the system can be updated again until
$n_-+n_0=0$.

\textit{Free Boundary conditions} ---
We will first describe our results for a system with free boundary conditions. The probabilities $\pi^c_h$, $\pi^c_v$
and $\pi^{c}_{hv}$  were already considered in equilibrium at $T_c$ by  Lapalme and Saint-Aubin  \cite{LSA}. These
authors measured these quantities on the triangular lattice.  They also tried to find a way to predict the behaviour of
these crossings in a way analogous to the Cardy formula for percolation. They obtained a differential equation on a four
point correlation function related to $\pi^c_h$ but were unable to solve it so they resorted to a numerical solution in
good agreement with their measurements. 

We considered systems of size $L \times L/r$ with $L=256$ while varying $r$. We only study the case $r \leq 1$, the case
with $r > 1$ being obtained by a trivial duality relation exchanging the vertical and horizontal directions. Our results
are shown in Fig.~\ref{PCI}. In this figure we compare the situation at $T_c$ and the final states of the evolution at
$T=0$. To do so we present the probabilities of getting a crossing $\pi^0_h, \pi^0_v$ and $\pi^0_{hv}$ for various
values of ratio $r$ obtained at $T=0$ as crosses and in the critical case, $\pi^c_h, \pi^c_v$ and $\pi^c_{hv}$ shown as
lines. The agreement between the $\pi^c$ and the $\pi^0$ is excellent, this is one of the main results of the present work. The values of the
$\pi^c$ are those obtained by Lapalme and Saint-Aubin in~\cite{LSA}.  Note that these authors considered the case of a
triangular lattice while our measurements have been done on a square lattice. As a check, we also repeated the
measurement of the $\pi^c$ on the square lattice and of the $\pi^0$ on the triangular lattice and our results are in
perfect agreement with the ones of Lapalme and Saint-Aubin obtained on the triangular lattice. This supports the
universality of these quantities. 

In conclusion, the crossing probabilities are the same for the equilibrium at $T_c$ and in the blocked states 
obtained after a quench at $T=0$ starting from the critical point. This first result confirms the idea that the final state is
dictated by the topology of spin clusters in the initial one.

 \begin{figure}[h!]
	 \begin{center}
\includegraphics[scale=0.2]{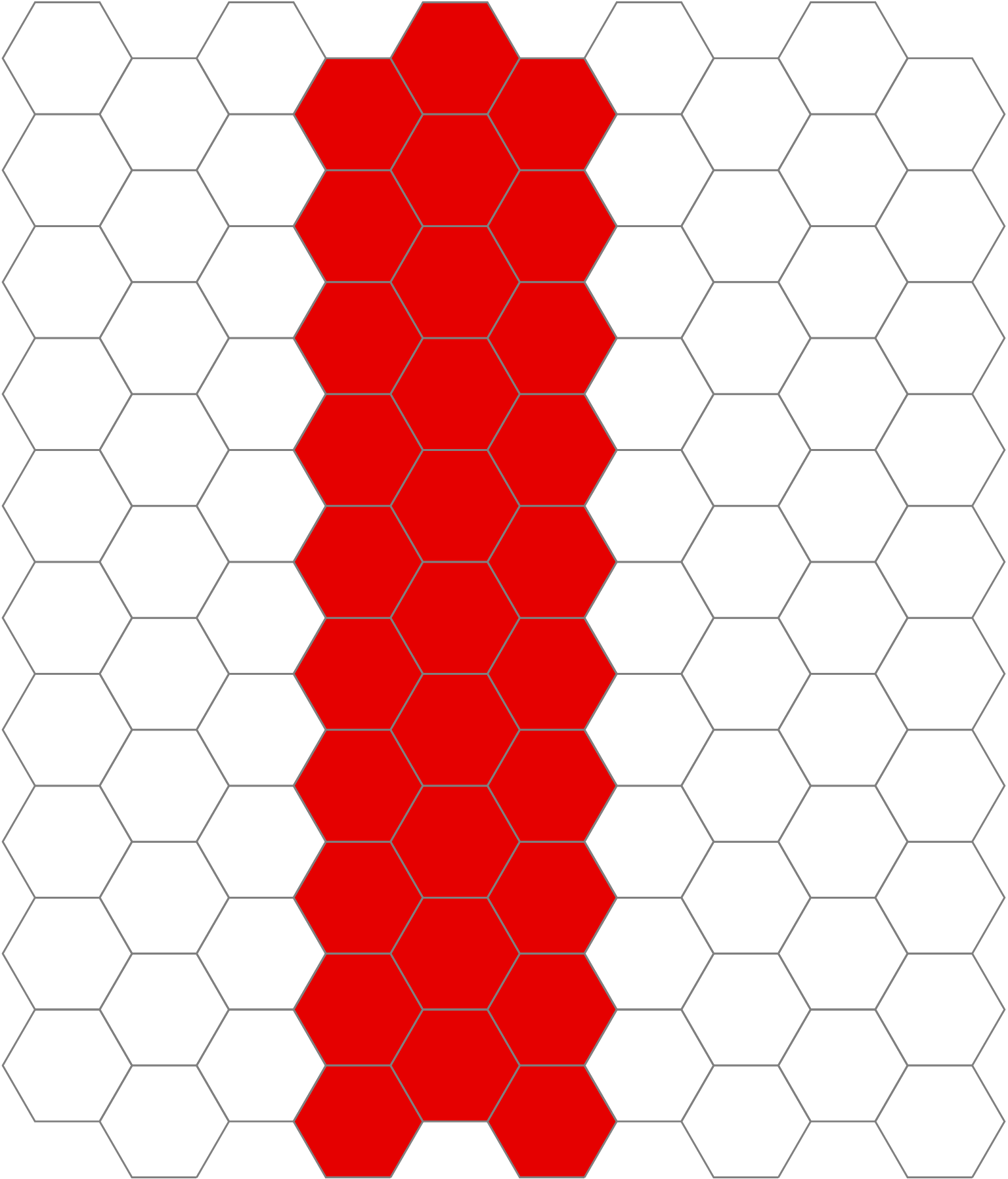}
\includegraphics[scale=0.2]{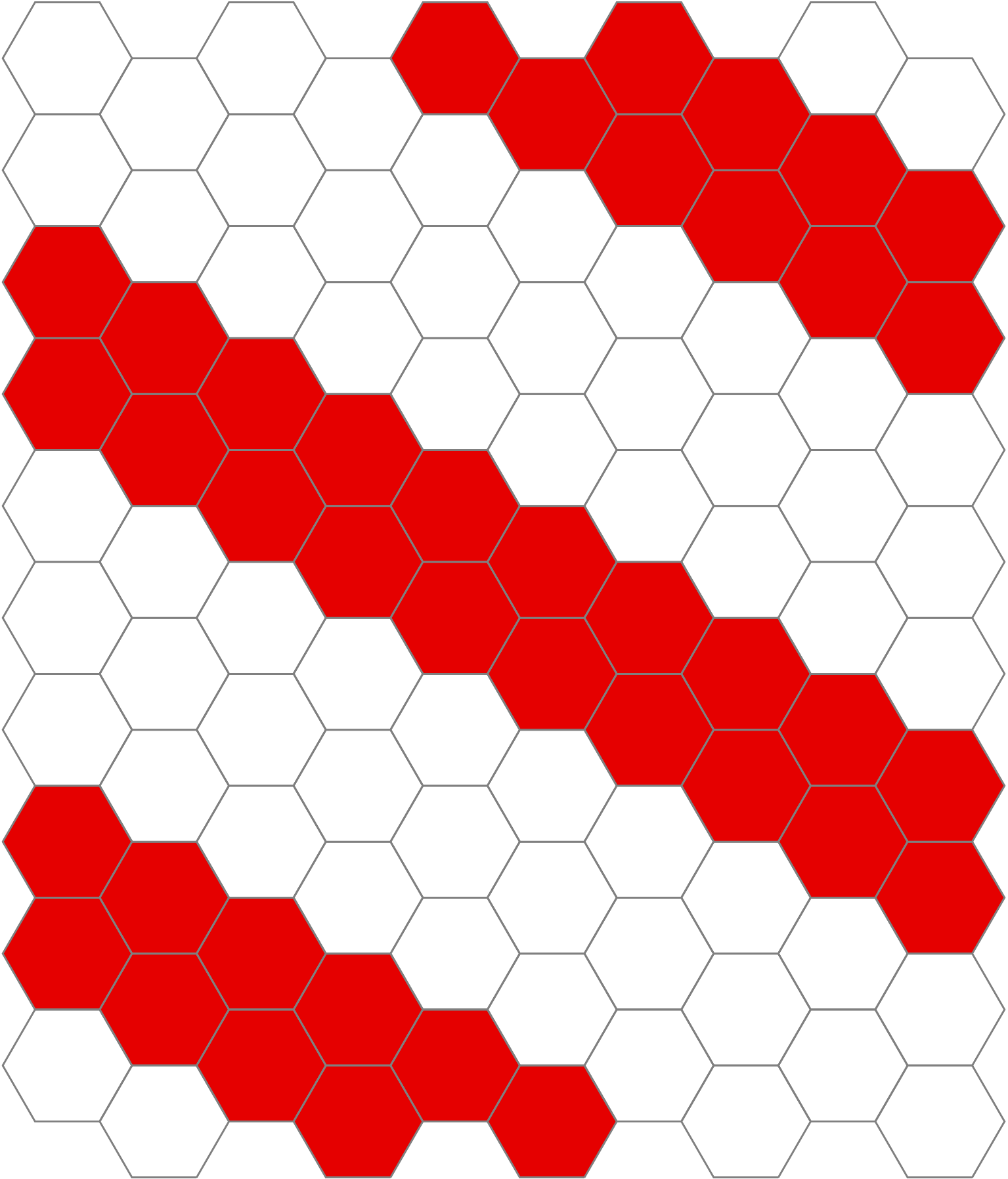}
	 \end{center}
 \caption{(color online) Two possible types of stripes of the alternate triangular lattice with periodic boundary
conditions. The vertical stripes on the left can exist for arbitrary aspect ratios $r$ contrarily to the stripes on the
right which only exists for $r=2n/\sqrt{3}$ with $n\in\mathbb{N}^*$. Actually the stripe on the right wraps in both directions
around the torus.}
 \label{sketch}
 \end{figure}

\textit{Periodic Boundary conditions} ---
To make this fact more explicit we move on to the case of periodic boundary condition (PBC). Here the situation is a bit
more complicated since clusters can wind in various ways around the torus. We can nonetheless define as previously the
percolation probabilities $\pi_h^c$, $\pi_v^c$, $\pi_{hv}^c$ and $\pi_h^0$, $\pi_v^0$, $\pi_{hv}^0$ of interest to us.
We consider the triangular lattice. With this choice of lattice and boundary condition only vertical stripes are stable
at $T=0$ for arbitrary aspects ratios (see Fig.~\ref{sketch}). This means that $\pi_h^0=0$ and $\pi_{hv}^0=1-\pi_v^0$
thus we only need to consider $\pi_v^0$. As is shown in Fig.~\ref{sketch}, diagonal stripes are stable in addition to
the vertical ones for the aspect ratios $r=2n/\sqrt{3}$ with $n\in\mathbb{N}^*$, but their rarity ($\pi_{\mathrm{diag}}$
is at most $6.10^{-8}$) forbids their study numerically as has been done in~\cite{olejarz_fate_2012}.
The factor $\sqrt3/2$ in the definition of the aspect ratio takes into account
the actual horizontal system size. We have seen that different percolation probabilities of same spin cluster for free
boundary conditions has been studied numerically and analytically in~\cite{LSA}. We found no such study for the PBC.
This motivated us to extend the work of Arguin~\cite{arguin_homology_2002} which deals with the probability that a given
configuration contains a Fortuin-Kasteleyn cluster winding $a$ times horizontally and $b$ times vertically around the
torus.  We have been able to obtain an explicit formula for such a probability for an Ising spin
cluster~\cite{blanchard_homology_2013}. In the case where the spin cluster winds only vertically around the torus, the
formula reduces to:
\begin{equation}
\pi^c_v(r)=\frac{1}{2r|\eta(ir)|}\frac{\theta_3\left(\frac{i}{12r}\right)-\theta_3\left(\frac{3i}{4r}\right)-2\theta_2\left(\frac{3i}{4r}\right)}{|\theta_2\left(ir\right)|+|\theta_3\left(ir\right)|+|\theta_4\left(ir\right)|}
\label{torus}
\end{equation}
where $\eta$ is the Dedekind $\eta$ function and the $\theta_i$ are the Jacobi $\theta$
functions.
\begin{figure}[h!]
	 \begin{center}
		\includegraphics[scale=1.15]{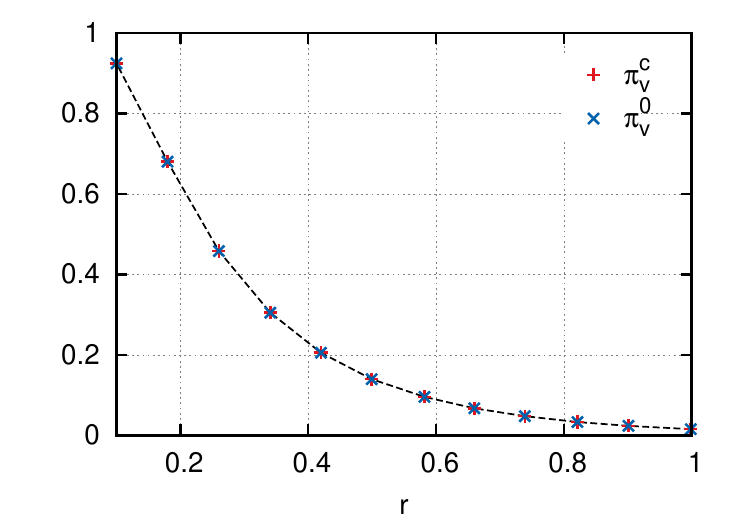}
	 \end{center}
 \caption{(color online) Vertical percolation probability $\pi_v^c$ at $T_c$ (red crosses) and vertical stripe probability $\pi_v^0$
 at $T=0$ (blue squares) for various aspect ratio $r=L_y/(\sqrt3L_x/2)$ with $L_y=128$. The dashed line corresponds to
the expression in Eq.~(\ref{torus}).}
 \label{pbc}
 \end{figure}
In Fig.~\ref{pbc}, the results for the triangular lattice with PBC are presented. The agreement
between $\pi^c_v$, $\pi^0_v$ (dots) obtained from the simulations and the theoretical prediction from Eq.~(\ref{torus})
for $\pi^c_v$ (dashed line) is very good. This second case confirms the link between probabilities of blocked stripe
states and initial crossing probabilities of spin clusters.

\textit{Fixed Boundary conditions} ---
Finally, we will look at a last case in which we impose fixed boundary conditions with $S=-1$ on the left and right sides
of a rectangle and with $S=+1$ on the top and bottom sides. This case is interesting because it has been extensively
studied analytically at $T_c$ by several authors using CFT and multiple stochastic Loewner
evolutions~\cite{arguin_non-unitary_2002,bauer_multiple_2005,kozdron_using_2009}, so there exists a number of
theoretical predictions to compare our simulations to; moreover it is really easy to simulate and analyze as we will see
below. 

\begin{figure}[ht!]
	\begin{center}
		\includegraphics[scale=2.0]{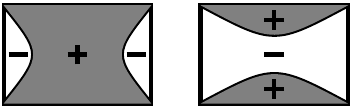}
	\end{center}
\caption{Schematic representation of the two configuration types with fixed boundary conditions. Is is easy to see that
beginning with the left (respectively right) configuration we will end up in a state where all spins equal $+1$ ($-1$)
except the fixed ones.}
\label{Fixed_BC}
\end{figure}

These boundary conditions will force the existence of two interfaces.  
The two possible configuration types are represented in Fig.~\ref{Fixed_BC}. 
In the first case, the first interface goes from the left top corner to the
left bottom corner and the second interface connects the right top corner to
the bottom right corner. In the second case the interface goes from the left
top corner to the right top corner and the second interface connects the bottom
left corner to the bottom right corner. The probabilities for these two
situations at $T_c$ can be written in several equivalent forms, we use
Kozdron's version written only in terms of the hypergeometric function
$F$~\cite{kozdron_using_2009}. The probability for the left situation in
Fig.~\ref{Fixed_BC} is given by:
\begin{equation}
P(x) = \frac{F(\frac{4}{3}, 3,\frac{8}{3};1-x)}{F(\frac{4}{3}, 3,\frac{8}{3};x)+ F(\frac{4}{3},3,\frac{8}{3};1-x)}
\end{equation} 
where $x\in(0,1)$ is related to the aspect ratio $r$ by:
\begin{equation}
r=\frac{K(x)}{K(1-x)}
\end{equation} 
with $K(x)$ the complete elliptic integral of the first kind. For the case of infinite initial temperature, the
probabilities for the two situations of Fig.~\ref{Fixed_BC} are given by Cardy formula~\cite{cardy_critical_1992}.
Indeed, in the case of percolation, the fixed boundary condition considered
in~\cite{arguin_non-unitary_2002,bauer_multiple_2005,kozdron_using_2009} coincides with the situation considered by
Cardy. 

\begin{figure}[h!]
	\begin{center}
		\includegraphics[scale=1.15]{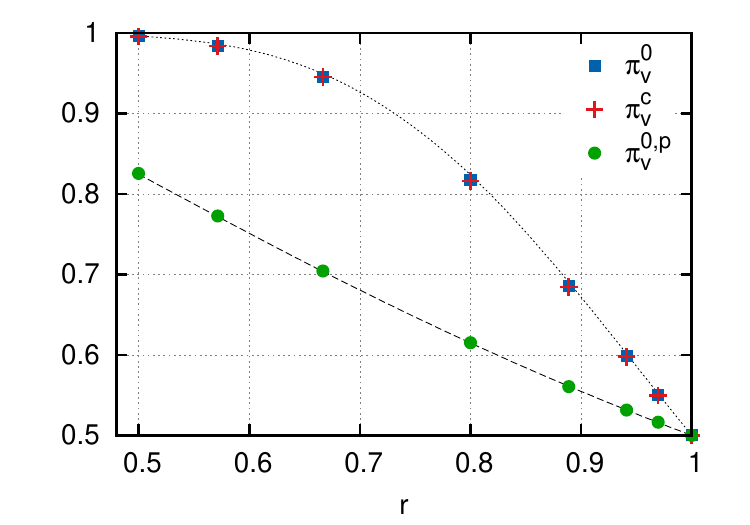}
	\end{center}
\caption{
(color online) Crossing probabilities versus $r$. The dashed lines correspond
to the formula proved
in~\cite{arguin_non-unitary_2002,bauer_multiple_2005,kozdron_using_2009}  for
the critical Ising model and in~\cite{cardy_critical_1992} for percolation. The
symbols correspond to our data for $\pi^0_v$ (blue filled square) (respectively
$\pi^{0,p}_v$ (green dots)) obtained after a quench at zero temperature from
$T_c$ (respectively $T=\infty$).  Note that the values measured for $\pi^c_v$
(red crosses) are indistinguishable from the probability $\pi^0_v$ since they
are numerically very close.}
\label{PC2}
\end{figure}

As previously we quench the system from $T_c$ to $T=0$ but also from $T=\infty$
to $T=0$ as we found no mention of this case in the literature. With this
boundary condition the analysis of the final state is really easy, the sign of
magnetization suffices to indicate the state of the system as bulk spins are
all of the same sign in the end. We show the results of our simulations for
$L=1280$ for the initial temperatures $T_c$ and $T=\infty$ in Fig.~\ref{PC2}.
The agreement with the theoretical prediction (dashed lines) is again excellent
in both cases.

{\it Conclusions} ---  
We have extended the connection between crossing probabilities and the probabilities of existence of stripe states after
a $T=0$ Glauber dynamics to the case of the critical Ising model for various boundary conditions. We have obtained clear
results showing that the final state of the evolution is strongly correlated to the initial condition, similarly to what
was already observed in~\cite{spirin_freezing_2001,spirin_fate_2001,barros_freezing_2009,olejarz_fate_2012} for the
quench from $T=\infty$ to $T=0$ in a similar setup. We expect that this is a general feature and it would be interesting
to check the generalisation of this result for other models. In a preliminary check, we have obtained similar results
for the $Q=3$ Potts model~\cite{blanchard_potts_2013}. We leave for the future similar studies for more complicated
models or even the case of larger dimensions.

{\bf ACKNOWLEDGMENTS}
We thank L.~F.~Cugliandolo and R.~Santachiara for useful discussions and L.~F.~Cugliandolo for her comments on the manuscript. 

\bibliography{stripes.bib}

\end{document}